\documentstyle[12pt,amssymb,epsfig]{article}
\author {Ernesto S. Loscar, Rodolfo A. Borzi and Ezequiel V. Albano$^{a}$\\
$^a${\it Instituto de Investigaciones Fisicoqu\'{\i}micas
Te\'{o}ricas y Aplicadas}\\{\it (INIFTA), UNLP, CONICET,
Suc.4, CC16,}\\{\it
1900 La Plata, Argentina}}

\title{Scaling behavior of jamming fluctuations upon
random sequential adsorption.}
\begin{document}
\maketitle

\begin{abstract}
It is shown that the fluctuations of the jamming coverage upon Random Sequential
Adsorption ($\sigma_{\theta_J}$), decay with the lattice size according
to the power-law
$\sigma_{\theta_J} \propto L^{-1/\nu_{J}}$, with
$\nu_{J} = \frac{2}{2D - d_{\rm f}}$,
where $D$ is the dimension of the substrate and $d_{\rm f}$ is the fractal dimension
of the set of sites belonging to the substrate where the RSA process actually takes
place.
This result is in excellent agreement with the figure recently 
reported by Vandewalle {\it et al} ({\it Eur. Phys. J.} B. {\bf 14}, 407 (2000)),
namely $\nu_{J} = 1.0 \pm 0.1$ for the RSA of needles with $D = 2$ and 
$d_{\rm f} = 2$, that gives $\nu_{J} = 1$. Furthermore, our prediction is in excellent
agreement with different previous numerical results. The derived relationships
are also confirmed by means of extensive numerical simulations applied to the
RSA of dimers on both stochastic and deterministic fractal substrates.
\end{abstract}

\newpage 

The irreversible deposition of particles on a surface involves
two characteristic time scales: the time between depositions,
and the diffusion time of the particles on the surface. For very strong
interaction between particles and the substrate (chemical adsorption),
diffusion becomes irrelevant and the Random Sequential Adsorption (RSA) 
model provides an excellent description of the underlying processes 
(for a review on RSA models see \cite{Evans}).
Under these conditions the system evolves rapidly toward far-from
equilibrium conditions and the dynamics becomes essentially
dominated by geometrical exclusion effects between particles.
This kind of effects has been observed in numerous experiments \cite{9}.

The RSA of needles (or linear segments)
on homogeneous, two-dimensional samples, has very recently attracted
considerable interest \cite{Galam,Pekalski}. Particular
attention has been drawn to the interplay between the jamming coverage 
and percolation \cite{Galam,Pekalski,frede}.
The percolation problem has also attracted considerable attention in the field of
statistical physics due to their relevance for the understanding of 
processes and phenomena in many other areas such as those occurring in 
disordered media, porous materials, systems of biological
and ecological interest, etc. \cite{hav1,hav2,stau}.
Therefore, a great progress in the field of the
statistical physics of far-from equilibrium processes
could be achieved by establishing links between
RSA and percolation \cite{Galam,Pekalski,frede}.

The percolation transition is related to the
probability of occurrence of an infinite connectivity between
randomly deposited objects, as a function of the fraction
$p$ of the substrate occupied by the objects.
Close to the percolation threshold $p_{c}$, 
the probability $P$ to find a percolating cluster, on a finite 
sample of side $L$, is given by an error function \cite{stau}
\begin{equation}
P = \frac{1}{\sqrt{2\pi}\sigma} \int_{-\infty}^{p}
exp\Big{[}-\frac{1}{2}
{\Big{(}\frac{p' - p_{c}}{\sigma}\Big{)}}^2
\Big{]}dp'  ,
\label{error}
\end{equation}
\noindent where $\sigma$ is the width of the transition
region. It is well known that $\sigma$ vanishes in the
thermodynamic limit according to \cite{stau}
\begin{equation}
\sigma \propto L^{-\frac{1}{\nu}},
\label{delta}
\end{equation}
\noindent where $\nu$ is the exponent
that governs the divergence of the correlation length
as $\xi \propto |p - p_{c}|^{-\nu}$.

Very recently it has been suggested that the jamming
probability and the fluctuations of the jamming coverage
may obey relationships similar to equations (\ref{error}) and
(\ref{delta}) \cite{Galam}, respectively. 
The aim of this note is to provide a qualitative derivation of equation 
(\ref{delta}) for the case of RSA on both homogeneous and deterministic 
fractal substrates. 
The predictions of the obtained equation will be compared with previously
published data and further numerical tests will be performed.
To accomplish these goals, the RSA of dimers on deterministic 
and stochastic fractals such as a Sierpinski Carpets (SC) \cite{hav2} 
and the diffusion front \cite{hav1,hav2}, has been studied. 

Let us first establish a link between the  fluctuations of the number 
of deposited particles (at the jamming state) on a subsystem of 
side $L_0$ with those of a system of side $L$, with $L_0 < L$. 
Considering a fractal subsystem of side $L_0$, that has itself $Q_0$ 
minimal pattern blocks, and increasing the size of the subsystem 
$n$ steps $\lambda$ times until reaching the size $L$, such 
as $L(n) = \lambda^n L_0$, then
$Q_0$ will change to $Q(n) = s^{n} Q_0$.  Therefore, eliminating $n$, 
it follows that 
\begin{equation}
Q={Q_0}\times \left({\frac{L}{L_0}} \right)^{d_{\rm f}}
\label{fract}
\end{equation}

\noindent where $d_{\rm f}=log(s)/log(\lambda)$ is the fractal dimension.

Let $N_0$ be the number of adsorbed particles in the starting 
subsystem of side $L_{0}$. For the system of side $L$ the number of
adsorbed particles $N(L)$ is given by the sum

\begin{equation}
N(L) = \sum_{\i=1}^{s^n} N_i ,
\label{sum1}
\end{equation}

\noindent where $N_i$ are the number of adsorbed particles on 
each subsystem of side $L_0$ that form the system of side $L$.  
Let $\sigma_{N(L_0)}$ be the fluctuations, in the starting 
$L_0$-subsystem, of $N_0$. If the correlation length associated to
the random sequential adsorption ($\xi_{Rsa}$) is short compared 
with $L_0$  ($\xi_{Rsa} << L_0$), the random variables $N_i$ will be 
statistically independent an so from Eq.(\ref{sum1})
it follows

\begin{equation}
\sigma^2_{N(L)}=
\sum_{\i=1}^{s^n} {\sigma_{i}}^2.
\label{sum2}
\end{equation}

\noindent Furthermore, in the thermodynamic limit, the $L_0$-subsystems 
should have identical statistical properties. Thus, their respective 
fluctuations will be the same. So, from the fact that 
$s^{n} = \frac{Q(n)}{Q_0}$ and 
Eqs.(\ref{fract}) and (\ref{sum2}) one has

\begin{equation}
\sigma^2_{N(L)}=\frac{\sigma^2_{N_0}}{L_0^{d_{\rm f}}} \times L^{d_{\rm f}},
\label{sum3}
\end{equation}

\noindent then the fluctuation of the density ($\theta$) in the
system of size $L$ can be obtained from Eq.(\ref{sum3})
dividing by $L^{2D}$, so that 

\begin{equation}
\sigma_{\theta} \propto  L^{-\frac{1}{\nu_{J}}} ,
\label{g141}
\end{equation}

\noindent where

\begin{equation}
\quad \nu_{J}= {\frac{2}{2D-d_{\rm f}}}. 
\label{g14}
\end{equation}

It should be stressed that Eqs. (\ref{g141}) and (\ref{g14}) are 
quite general relationships valid for substrate systems that are both 
homogeneous and deterministic fractal. Furthermore, the same relationships
hold for the case of substrates globally-invariant under translations, 
such as random fractals, as it has been demonstrated elsewhere \cite{nosot}. 
Also, the condition that the correlation 
length of the RSA process should be smaller than the system size is usually 
valid for jammed states, where the correlation length is very short.
It is also very interesting to notice that, using these relationships it may 
be possible to evaluate $d_{\rm f}$ performing both RSA numerical simulations and 
actual experiments. Furthermore, existing numerical simulations
performed in $D = 2$ dimensions with $d_{\rm f} = 2$ are
in excellent agreement with equations (\ref{g141}) and (\ref{g14})
(notice that for these conditions it follows straightforwardly from
equation (\ref{g14}) that $\nu_{J} = 1$ exactly). In fact,
for the jamming upon RSA of needles in two dimensions the value
$\nu_{J} = 1.0 \pm 0.1$ has been reported \cite{Galam} and this
figure is independent of the aspect ratio of the needles.
Furthermore, early numerical results of Nakamura for the 
RSA of square blocks are also consistent with $\nu_{J} \simeq 1$ \cite{Naka}, 
while Kondrat {\it et al.} \cite{Pekalski} have
reported $\nu_{J} = 1.00 \pm 0.05$ for the RSA of 
segments on the square lattice. Since the obtained values for the
exponent are independent (within error bars) of: i) the
length of the segments (for all a = 1,2,....,45) \cite{Pekalski},
ii) the aspect ratio of the needles \cite{Galam} and iii) 
the size of the square blocks \cite{Naka},
it has been suggested that $\nu_{J}$ is a good candidate for 
an universal quantity of the jamming process \cite{Pekalski}.
Within this context, our finding shows that $\nu_J$ depends on the
dimensionality of the substrate and the set where the RSA processes 
actually takes place.

On homogeneous samples the jamming coverage $(\theta$) and its
fluctuations ($\sigma_{\theta}$) can straightforwardly be obtained, since
one has to deal with a single stochastic process. However,
RSA on nonhomogeneous random substrates 
requires a careful treatment because 
two correlated stochastic processes are now involved \cite{nosot}.
One can assume that the fluctuations due to the RSA process
are given by an average over $M$ independent samples:

\begin{equation}
\sigma_{\theta}=
\sum_{i=1}^{M} 
\frac { {\sigma^{i}_{\theta}}}
{M} ,
\label{sigmarsa}
\end{equation}

\noindent where $\sigma^{i}_{\theta}$ are the fluctuations measured
using a {\bf single} substrate sample but taken averages over 
independent RSA trials.
It has been shown \cite{nosot} that measuring  $\sigma_{\theta}$ with the 
aid of Eq. (\ref{sigmarsa}) one captures the physical behavior of 
the RSA process. In contrast, measuring the fluctuations
of the average jamming coverage of different samples
the physical behavior reflects the properties of the substrate \cite{nosot}.  

In order to perform additional tests to the obtained analytical results, 
the RSA of dimers on both stochastic and deterministic fractals
has been studied numerically.

As example of an stochastic fractal, we have used a diffusion front.
In order to generate the diffusion front, we considered the diffusion of 
particles at random, but with hard-core
interactions, on a 2D square lattice of size $L \times L$.
There is a source of particles at the first row of the lattice 
$y = 1, 1 \leq x \leq L$ kept at concentration $p(1,t) \equiv 1$.
Also, at row $y = L+1, 1 \leq x \leq L$ there is a well,
$p(L+1,t) \equiv 0$. So, there is a concentration gradient
along the source-well direction, while along
the perpendicular {\it x}-direction periodic boundary conditions are
imposed.
In the steady state the concentration gradient is constant, so one has

\begin{equation}
\nabla p(y)=L^{-1}.
\label{grad}
\end{equation}

It is well known that the properties of the diffusion front \cite{SRG,MAR1}
are closely related to those of the incipient percolation 
cluster \cite{hav1,hav2,stau}. 
As the concentration $p(y)$ of particles depends on the position,
decreasing from the source to the well, one actually has a
{\bf gradient percolation} system. The
structure of the diffusion front is identical to the structure of
the hull of the incipient percolation cluster \cite{SRG}. Furthermore,
the concentration of particles at the mean front position $y_f$ is the
same as the percolation threshold $p_c$, so that $p(y_f)=p_c$ \cite{SRG}.
The diffusion front is conveniently described by its
average width $\sigma_f$ and the total number $N_f$ of particles
that constitute it.
Using heuristic arguments it has been suggested that 
\begin{eqnarray}
\frac {N_f}{L}\sim \left| \nabla p(y_f) \right|^{-\alpha_N} \qquad \qquad \textrm{where}\
\quad \alpha_{N} = \frac{1}{\nu+1} 
\label{alfa} 
\end{eqnarray} 
being $\nu$ the critical exponent of the correlation length in the percolation
problem \cite{hav1,hav2}; $\nu = 4/3$ in $2D$, which gives  $\alpha_N=3/7$.
So, from Eqs.(\ref{grad}) and (\ref{alfa}) one has     

\begin{equation}
N_f \sim L^{d_{\rm f}^{DF}}  ,
\label{frente}
\end{equation}

\noindent with $d_{\rm f}^{DF} = \alpha_{N} + 1 = 10/7 \approx 1.4286$,
and the diffusion front is a stochastic self-similar fractal \cite{note}.

RSA of dimers on diffusion fronts has been simulated using two rules: 
according to {\bf Rule I} only adsorption events of dimers taking place 
on two nearest-neighbor (NN) sites, such us one of then belongs 
to the diffusion front  and the remaining one is outside it, 
are considered. On the other hand, using {\bf Rule II} one only allows 
the adsorption on NN sites of the diffusion front, disregarding adsorption 
trials on already occupied sites of the front and sites outside the fractal.  

RSA of dimers on deterministic fractals (Sierpinski 
Carpets \cite{hav1,hav2}) is
also studied.
The SC in $D = 2$ dimensions is generated by dividing a full square 
into $\lambda^D$  smaller squares of the same size. Out of these squares,
$k$ of them are chosen and removed. In the next iteration, the procedure 
is repeated by dividing each of the small squares left into $\lambda^D$
smaller squares removing those $k$ squares that are located at the 
same positions as in the first iteration.
The resultant fractal dimensions are
\begin{equation}
d_{\rm f}(s,\lambda) = log(s)/log(\lambda) 
\end{equation}
where $s=\lambda^D - k$. In principle, this procedure has to be
repeated again and again, however for the practical implementation in 
a computer only a finite number of iterations are actually 
performed \cite{hav2,nazareno}.
In a square lattice the smaller subdivision is actually a single site and the
length is measured in site units. Furthermore there is a minimal pattern of $\lambda^{d_{\rm f}}$ sites.
In the present work various generations of SC's 
of different size $L$, with periodical boundary conditions,
have been employed. In all cases dimers are allow to adsorb only 
on NN empty sites belonging to the fractal. 
For SC's with $\lambda = 3$ and $k = 1,2,3$,
as used in the simulations, the fractal dimensions are 
$ d_{\rm f_{I}}=log(8)/log(3)\approx 1.8928$, 
$ d_{\rm f_{II}}=log(7)/log(3)\approx 1.7712$ and 
$ d_{\rm f_{III}}=log(6)/log(3)\approx 1.6309$,
respectively.

\begin{figure}
\centerline{\epsfysize=9.0 cm \epsffile{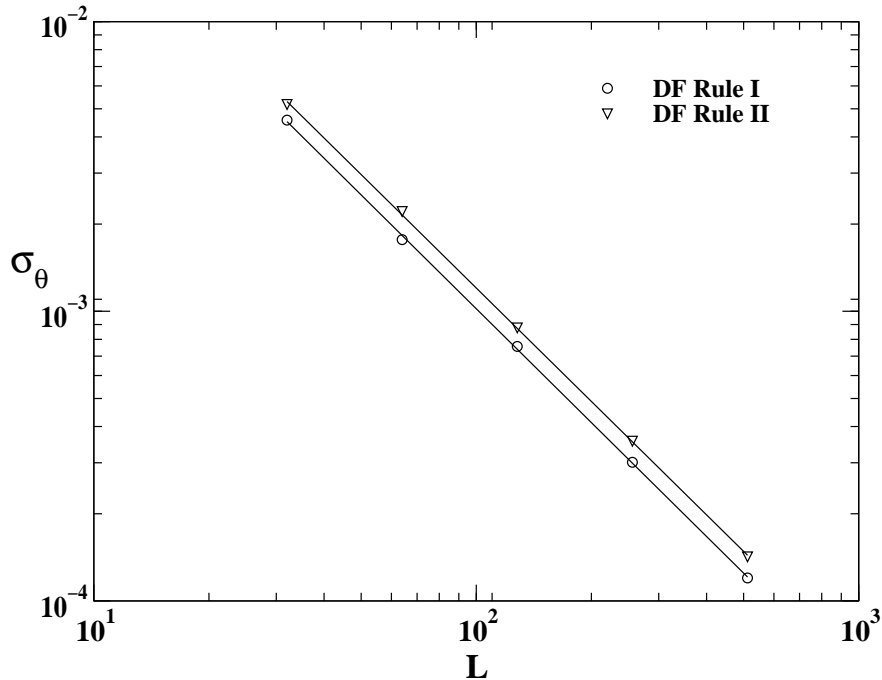}}
\caption {Log-log plots of $\sigma_{\theta}$ versus $L$ for the case of RSA of dimers
on the random fractal generated by diffusion fronts (DF). Results obtained using two
different adsorption rules are shown. For details the adsorption rules see the
text.}
\label{randomfractals}
\end{figure}

\begin{figure}
\centerline{\epsfysize=9.0 cm \epsffile{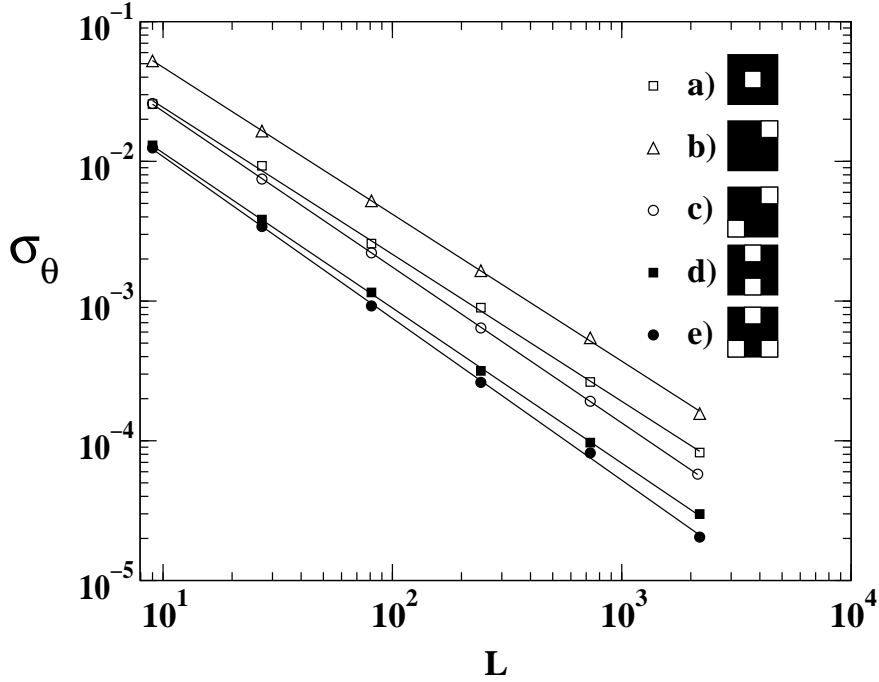}}

\caption {Log-log plots of $\sigma_{\theta}$ versus $L$ for the case of 
RSA on Sierpinski Carpets obtained using different generating patterns as 
shown in the figure (black squares compose the fractal structure).}
\label{sierpi}
\end{figure}

Figures \ref{randomfractals} and \ref{sierpi} 
show log-log plots of $\sigma_{\theta}$ versus $L$
obtained upon RSA of dimers on diffusion fronts and Sierpinski Carpets, 
respectively.  The obtained  results, for these kind of fractals, are in 
excellent agreement with the prediction of Eq.(\ref{g14}) as follows
from the comparison of evaluated and theoretical exponents listed
in Table I. Further support to the theoretical prediction follows from
additional results obtained using homogeneous samples, which are also listed
in Table I.


\begin{table}
\caption{Examples of the application of Eq.(\ref{g14}) to different 
fractals as listed in the first column: SC $\equiv$ Sierpinski Carpet,
DF $\equiv$ Diffusion front, HS2 Homogeneous Substrate in $D = 2$ 
dimensions. The 2nd column shows 
the exponents obtained fitting Eq.(\ref{g14}) to the simulation results 
while the 3rd one shows the estimations of $d_{\rm f}$ obtained using 
$\frac{1}{\nu_{J}}= {\frac{2D-d_{\rm f}}{2}}$.
The 4th column is a list of the exact values of $d_{\rm f}$.
Notice that for SC the labels a)-e) allows to identify the generating 
patterns, as shown in figure \ref{sierpi}. }
\begin{center}
\begin{tabular}{|c|c|c|c|}
\hline
 Substrate   & $1/\nu_{J}$& $d_{\rm f}^*$&  $d_{\rm f}$                   \\ \hline 
 SC  (a)     & $1.051(4)$ & $1.898(8)$  &   $ln(8)/ln(3) \simeq 1.893$    \\ \hline 
 SC  (b)     & $1.052(4)$ & $1.896(8)$  &   $ln(8)/ln(3) \simeq 1.893$    \\ \hline
 SC  (c)     & $1.115(2)$ & $1.770(4)$  &   $ln(7)/ln(3) \simeq 1.771$    \\ \hline
 SC  (d)     & $1.110(7)$ & $1.780(15)$ &   $ln(7)/ln(3) \simeq 1.771$    \\ \hline
 SC  (e)     & $1.16(2)$  & $1.68(4)$   &   $ln(6)/ln(3) \simeq 1.631$    \\ \hline
 DF          & $1.30(2)$  & $1.40(4)$   &   $10/7        \simeq 1.429$    \\ \hline
 HS2 (D = 2) & $1$        & -           &   2                             \\ \hline 
\end{tabular}
\end{center}   
\end{table}

Summing up, it is shown that the exponent
$\nu_{J}$ can be obtained as a function of the 
dimensionality $D$ of the space and the 
fractal dimension $d_{\rm f}$ of the
subset site where the RSA process actually takes place.
Our main result $\nu_{J}= {\frac{2}{2D-d_{\rm f}}}$,
provides a solid ground to previous numerical
data \cite{Galam,Pekalski,Naka}. 
Furthermore, in this work, the validity of the proposed relationship 
is verified by means of extensive numerical simulations, 
using both homogeneous substrates as well as different  
fractals. 

{\bf Acknowledgments}: This work was supported by CONICET, 
UNLP and ANPCyT (Argentina).



\begin{thebibliography}{99}
\bibitem{note} Notice that scaling $N_{f}$ as a function of the
diffusion width, instead of the lattice size $L$ as 
in Eq.(\ref{frente}), one can define a different fractal dimension given by
$d_{f}^{*DF} = \nu +1 / \nu = 7/4$.
\end{thebibliography}
\end{document}